\documentclass[aps,prc,twocolumn,superscriptaddress,floatfix,longbibliography]{revtex4-2}

\usepackage{amsmath,amssymb,bm}
\usepackage{graphicx}
\usepackage{booktabs}
\usepackage{xcolor}
\usepackage[version=3]{mhchem}
\usepackage{hyperref}
\hypersetup{colorlinks=true,linkcolor=blue,citecolor=blue,urlcolor=blue}

\def\be{\begin{eqnarray}}
\def\ee{\end{eqnarray}}

\newcommand{\Chat}{\hat{C}}
\newcommand{\Cl}{\Chat_\ell}
\newcommand{\kC}{\ensuremath{k_{\mathrm C}}}
\newcommand{\Zeff}{\ensuremath{Z_{\mathrm{eff}}}}
\newcommand{\Gm}{\ensuremath{\Gamma(2+\eta_B)}}
\newcommand{\Dcal}{\mathcal{D}}

\newcommand{\Csq}{\ensuremath{C^2}}
\newcommand{\Cp}{\ensuremath{{}^{12}\mathrm{C}}}
\newcommand{\Nt}{\ensuremath{{}^{13}\mathrm{N}}}
\newcommand{\Etwo}{\ensuremath{E_{2^+}}}

\begin{document}
\raggedbottom

\title{Radiative proton capture on ${}^{12}\mathrm{C}$ in cluster effective field theory}

\author{Tae-Sun Park}
\email{tspark@ibs.re.kr}
\affiliation{Center for Exotic Nuclei Studies, Institute for Basic Science, Daejeon 34126, Korea}

\author{Eunjin In}
\affiliation{Department of Physics and Astronomy, Louisiana State University,
Baton Rouge, Louisiana 70803, USA}

\author{Young-Ho Song}
\affiliation{Institute for Rare Isotope Science, Institute for Basic Science, Daejeon 34000, Korea}

\begin{abstract}
Radiative proton capture on ${}^{12}\mathrm{C}$ to the ground state of
${}^{13}\mathrm{N}$ is calculated in cluster effective field theory through
next-to-leading order. At low energy the reaction is dominated by
$s_{1/2}\to p_{1/2}$ $E1$ capture, which we treat in the long-wavelength
limit. Its amplitude is strongly hindered by destructive interference
between the contribution fixed by gauge invariance and a counter-term $E1$
current, making the capture particularly sensitive to subleading terms. We
fit the recent data below $0.95$~MeV, determining the $E1$ transition
counter-term coefficients together with the ground-state asymptotic
normalization coefficient and the $\frac{1}{2}^+$ resonance and shape
parameters. We obtain $S(0)=1.34\pm0.07$~keV\,b and
$S(25~\mathrm{keV})=1.43\pm0.07$~keV\,b, the latter in agreement with recent
$R$-matrix extrapolations. We also estimate the effects of the omitted
$p$-wave $M1$ capture and $\frac{3}{2}^-$ resonance tail, and find that the
fitted asymptotic normalization coefficient moves substantially while the
$S$ factors remain almost unchanged.
\end{abstract}

\keywords{cluster effective field theory, radiative capture, CNO cycle,
astrophysical $S$ factor, asymptotic normalization coefficient}

\maketitle

\section{Introduction}

The reaction $\Cp(p,\gamma)\Nt$ is the first proton capture in the CNO cycle.
At stellar energies its cross section is too small to be measured directly,
so the reaction rate must be extrapolated from measurements at higher
energies. The extrapolation is strongly influenced by the
$J^\pi=\tfrac12^+$ state of $\Nt$ at $E=424.6$~keV in the center-of-mass
frame, since stellar energies lie on its low-energy tail.

The CNO cycle supplies about one percent of the Sun's energy, and its
neutrino flux, measured by Borexino~\cite{borexino2020}, constrains the solar
carbon and nitrogen abundances~\cite{solarfusion3}. The
$\Cp(p,\gamma)\Nt$ rate also affects the $^{12}$C/$^{13}$C ratio during
non-equilibrium CN burning. Both applications require the reaction rate at
energies of roughly $20$--$100$~keV.

Experimental information has improved considerably in recent years. LUNA and
the Felsenkeller group measured the resonance region and its low-energy
tail~\cite{luna2023,skowronski2023}, Kettner \emph{et al.} provided an
independent absolute measurement of the cross section and resonance
parameters~\cite{kettner2023}, and Gy\"urky \emph{et al.} extended the
activation data over a wider energy range~\cite{gyurky2023}. Nevertheless,
a recent comparison of $R$-matrix analyses found a sensitivity of the
low-energy extrapolation to the fitting procedure~\cite{skowronski2025},
and the 2025 evaluation of experimental thermonuclear reaction rates
(ETR25)~\cite{etr25} did not recommend a rate for this reaction. This
situation motivates an analysis in a framework with a systematic
low-energy expansion.

Cluster effective field theory (EFT) treats the relevant nuclei as explicit
degrees of freedom and has been applied to a variety of low-energy
radiative-capture reactions
\cite{zhang2015,zhang2018,higa2018,premarathna2020,khadka2025,
Ryb2014,RYBERG2016,fernando2012,son2022,ando2019capture}.
For the present system, elastic $p$-$\Cp$ scattering below
$E_p=2$~MeV was analyzed in the same framework~\cite{in2024}, including the
$s_{1/2}$, $p_{3/2}$, and $d_{5/2}$ resonances of $\Nt$, and in one of its three
fits the $\tfrac12^-$ ground state as well. Here we extend the
cluster-EFT description to radiative capture into the
$\tfrac12^-$ ground state. Capture to the first $\tfrac12^+$ state has been
studied in halo EFT~\cite{khansari2017}. A similar framework
has been used for $E1$ capture in
$\Cp(\alpha,\gamma)^{16}$O~\cite{ando2019capture} and for radiative proton
capture on $^{15}$N~\cite{son2022}.

At stellar energies the ground-state transition is dominated by
$s_{1/2}\to p_{1/2}$ $E1$ capture. Direct $d_{3/2}\to p_{1/2}$ capture
becomes increasingly important toward the upper end of the measured energy
range, while $p$-wave $M1$ capture and higher multipoles also begin to
contribute. In the present calculation we retain the $E1$ contributions in
the long-wavelength limit, and we work to next-to-leading order (NLO) in the
expansion described in Sec.~\ref{sec:counting}.

A distinctive feature of this reaction is that the $s$-wave $E1$ amplitude
is strongly hindered. The contribution fixed by gauge invariance and the
counter-term $E1$ contribution are nearly equal in magnitude and
opposite in sign, leaving a 
net amplitude about an order of magnitude
smaller than either term separately. The cancellation makes the
capture amplitude particularly sensitive to the strong and electromagnetic
low-energy constants. We therefore determine the $s_{1/2}$ resonance
parameters together with the capture amplitude. The strong-interaction
parameters of Ref.~\cite{in2024} are used for comparison and are not imposed
on the capture fit.

The $p_{1/2}$ ground-state pole is fixed by the proton separation energy and
the squared asymptotic normalization coefficient (ANC) $\Csq$. We determine
$\Csq$ from the capture data at each order, and then compare it with values
extracted from transfer reactions~\cite{li2010,artemov2022}. At NLO
we include the leading energy dependence of the
counter-term $E1$ current
and the $\tfrac12^+$ effective-range shape term. We also evaluate the shift
generated at next-to-next-to-leading order and use it to estimate the
truncation uncertainty.

Our adopted NLO result is
$S(0)=1.34\pm0.07$~keV\,b and
$S(25~\mathrm{keV})=1.43\pm0.07$~keV\,b, the latter consistent with the
recent $R$-matrix extrapolations of Refs.~\cite{kettner2023,skowronski2023}.

Section~\ref{sec:amplitude} gives the $E1$ amplitude and the cancellation it
contains, Sec.~\ref{sec:input} fixes its parameters, and
Sec.~\ref{sec:results} presents the fit and the uncertainty budget.

\section{Framework}
\label{sec:framework}

\subsection{Scales and power counting}
\label{sec:counting}

The 
cluster
EFT treats the proton and the $\Cp$ nucleus as explicit low-energy
degrees of freedom and organizes observables in powers of $Q/\Lambda$, where $Q$
denotes the low-momentum scales of the process and $\Lambda$ the scale at
which degrees of freedom omitted from the EFT become resolved.

The scales used below are collected in Table~\ref{tab:scales}. The reduced mass is
$\mu=m_pm_c/M=865.594$~MeV with 
total mass $M=m_p+m_c$, and the Coulomb momentum is
$\kC=Z_p Z_c\alpha\mu$ with 
$Z_p=1$, $Z_c=6$ and $\alpha=1/137.036$.
For the $p_{1/2}$ ground state of $\Nt$, 
the proton separation energy is
$S_p=1943.49(27)$~keV~\cite{ame2020}, the binding momentum is
$\gamma=\sqrt{2\mu S_p}$, and the Sommerfeld parameter is
$\eta_B=\kC/\gamma=0.65338$.
The $2^+$ excitation of $\Cp$ at $\Etwo=4.44$~MeV sets the breakdown scale,
with the corresponding momentum $\Lambda=\sqrt{2\mu\Etwo}$.

The characteristic low scales are $\kC$, $\gamma$, the resonance momentum
$k_R$, and the relative momentum $p$. Their ratios to $\Lambda$ range from
about $0.3$ to $0.66$, while $p/\Lambda$ decreases to $0.075$ at
$25$~keV. The system is therefore not a halo system with a parametrically
small binding momentum, since $1/\gamma=3.40$~fm is only $1.4$ times the
$\Cp$ charge radius. The binding momentum enters the final-state wave
function exactly rather than through the $Q/\Lambda$ expansion, and the
order-by-order corrections of Sec.~\ref{sec:orders} are controlled by
$k^2/\Lambda^2=E/\Etwo$.
The usefulness of the cluster description for the capture
amplitude is nevertheless supported by its peripheral character. At
$25$~keV, $96\%$ of the external $E1$ radial integral comes from
$r>1/\Lambda$ (Appendix~\ref{app:e1}).

\begin{table}[t]
\centering
\caption{Momentum scales. Here $\Etwo=4.44$~MeV is the $\Cp(2^+)$
excitation energy and $\Lambda=\sqrt{2\mu\Etwo}$.}
\label{tab:scales}
\begin{tabular}{lccc}
\toprule
 & MeV & fm$^{-1}$ & ratio to $\Lambda$ \\
\midrule
$\kC=Z_pZ_c\alpha\mu$      & 37.90 & 0.1921 & 0.432 \\
$\gamma=\sqrt{2\mu S_p}$    & 58.00 & 0.2940 & 0.662 \\
$k_R$ ($\tfrac12^+$ resonance) & 27.11 & 0.1374 & 0.309 \\
$p$ at $E=25$~keV           &  6.58 & 0.0333 & 0.075 \\
$p$ at $E=400$~keV          & 26.31 & 0.1334 & 0.300 \\
\midrule
$\Lambda=\sqrt{2\mu\Etwo}$   & 87.67 & 0.4443 & 1 \\
\bottomrule
\end{tabular}
\end{table}

In the strong sector the narrow $\tfrac12^+$ resonance is described at
leading order (LO) by two effective-range parameters~\cite{gelman2009,in2024}, which we take as
the expansion point $k_r^2$ and the slope $r'_0$ of
Eq.~(\ref{eq:pole-ere}). Equivalently, they may be
fixed by the resonance energy $E_R$ and proton width $\Gamma_p$
(Sec.~\ref{sec:swave}). The
$p_{1/2}$ bound state also requires two LO constants
~\cite{bertulani2002,Ryb2014,zhang2018}, fixed here by the separation energy
$S_p$ and the ANC $C$ (Sec.~\ref{sec:anc}).

In the electromagnetic sector, the one-body currents and the seagulls fixed
by gauge invariance are parameter free. A short-distance $E1$ operator joins
the initial- and final-state dicluster fields. For a proton halo the
initial-state strong interaction is dropped, since the Coulomb barrier
suppresses it exponentially at threshold, and the capture is then free of
short-distance counterterms at LO~\cite{Ryb2014}. Here the entrance channel
carries the narrow $\tfrac12^+$ resonance, so that interaction has to be
kept. It brings the divergence (Appendix~\ref{app:e1}), which we renormalize
with the counter-term $E1$ operator. With $\gamma/\Lambda=0.662$ there is no
parametric reason to expect its contribution to be small. The counter-term
therefore enters at LO. Section~\ref{sec:hindrance} shows its numerical
importance in the present reaction.

\subsection{Lagrangian}

We use the effective Lagrangian of Ref.~\cite{in2024}, supplemented by
the electromagnetic operators needed for capture,
\begin{align}
\mathcal{L}&=
\psi_p^{\dag}\left(iD_t+\frac{\bm D^2}{2m_p}\right)\psi_p
+\psi_c^{\dag}\left(iD_t+\frac{\bm D^2}{2m_c}\right)\psi_c
\nonumber\\
&\quad+\sum_x d_x^{\dag}\left[\Delta_x+\sum_{n=1}^{N_x}\nu_{n,x}
\left(iD_t+\frac{\bm D^2}{2M}\right)^n\right]d_x
\nonumber\\
&\quad-\sum_x g_x\left[d_x^{\dag}\,\mathcal{P}_x(\psi_p,\psi_c)
+\mathrm{h.c.}\right]
+
\mathcal{L}_{E1}
+\cdots,
\label{eq:lag}
\end{align}
Here $\psi_p$ and $\psi_c$ denote the proton and $\Cp$ fields, and $d_x$
the dicluster field in channel $x=J^\pi$. Its parameters are the residual mass
$\Delta_x$, the kinetic coefficients $\nu_{n,x}$, and the coupling $g_x$;
$N_x$ specifies the number of effective-range terms retained in that channel.
$D_\mu=\partial_\mu+ie\hat Q A_\mu$, with $\hat Q$ the charge operator in
units of $e$.
The operator $\mathcal{P}_x$ contains
zero, one, or two derivatives for the $s$, $p$, or $d$ wave, respectively.
Its structure for
$x=\tfrac12^+,\tfrac32^-,\tfrac12^-,\tfrac52^+$ is given in
Appendix~\ref{app:spin}.

The leading counter-term $E1$ operator is
\begin{equation}
\mathcal{L}_{E1}= i e\,\mathring c\;d^{\dag}_{\frac12^-,s'}\,
C^{\frac12 s'}_{1\alpha,\frac12 s}\,E^\alpha\,d_{\frac12^+,s}
+\mathrm{h.c.},
\label{eq:LE1}
\end{equation}
where $E^\alpha$ is a spherical component of the electric field and
$\mathring c$ is the corresponding low-energy constant. We mark bare
couplings with a circle.

For the $s_{1/2}\to p_{1/2}$ $E1$ transition, the higher-order
short-distance contributions can be represented by promoting $\mathring c$ to
an energy-dependent strength
$\mathring c(E)=\sum_{n\ge0}\mathring c_nE^n$.
One of the roles of these counter-terms is to absorb the singularity
appearing in the loop diagrams. To this end we write
\begin{equation}
\mathring c(E)=c_{\rm div}(E)+c(E),
\label{eq:cren}
\end{equation}
with
\begin{equation}
c_{\rm div}(E)=\frac{\mu}{3\pi\,\omega}
\left(\frac{1}{2r_{\min}}-2\kC\right),
\label{eq:cdiv}
\end{equation}
where $\omega=E+S_p$ is the photon energy and $r_{\min}$ the radial regulator
of Appendix~\ref{app:e1}, taken to zero at the end. The form and the overall
factor are chosen to cancel $\mathcal A_{\rm div}$ of
Eq.~(\ref{eq:diagrams}) completely. We are then left with $c(E)$ alone,
\begin{equation}
c(E)=\sum_{n\ge0}c_nE^n=c_0+c_1E+c_2E^2+\cdots ,
\label{eq:cE}
\end{equation}
where the $1/\omega$ of Eq.~(\ref{eq:cdiv}) is analytic at threshold because
$S_p>0$ and is absorbed order by order into the $c_n$.

The naive expansion parameter here is $E/\Etwo=k^2/\Lambda^2$, and the same
factor suppresses each successive shape term in the effective-range
function. The observed convergence to NLO is faster than this
estimate, as shown in Sec.~\ref{sec:cv}. The low-energy constants that enter
at NLO are listed in Sec.~\ref{sec:orders}.

\section{The $E1$ capture amplitude}
\label{sec:amplitude}

The capture 
is dominated by the $s_{1/2}\to p_{1/2}$ $E1$ transition at low energies.
We also include direct
$d_{3/2}\to p_{1/2}$ $E1$ capture, which introduces no additional
low-energy constants and contributes incoherently to the
cross section (Appendix~\ref{app:other}). Its fraction of the total $S$
factor increases from $0.9\%$ at 25~keV to $61\%$ at 0.92~MeV because the
$s$-wave amplitude falls toward a zero near 1.4~MeV. The $\tfrac52^+$ entrance channel
cannot connect to the ground state by $E1$ radiation. We neglect higher multipoles
and $p$-wave $M1$ capture, whose contribution is estimated in
Appendix~\ref{app:other}.

For the $\tfrac12^+\to\tfrac12^-$ transition, angular momentum and parity
allow only the $E1$ multipole. The transition therefore has a single reduced
amplitude $\mathcal A(E)$,
\be
iT_{fi}=ie\, \bm\epsilon^{*}(\bm k_\gamma,\lambda)\cdot
\bm J_{fi},\qquad
\bm J_{fi}=\chi_x^\dagger\,\mathcal A(E)\,\bm\sigma\,\chi_p.
\ee
With $p=\sqrt{2\mu E}$, its cross section and the $S$
factor are
\begin{equation}
\sigma=\frac{4\alpha\omega\mu|\mathcal A|^2}{p},
\qquad S(E)=\sigma E e^{2\pi\eta}.
\label{eq:xsec}
\end{equation}
The $d_{3/2}$ contribution is added to $\sigma$ incoherently.

\subsection{Diagrams and the long-wavelength amplitude}

\begin{figure*}[t]
\centering
\includegraphics[width=0.92\textwidth]{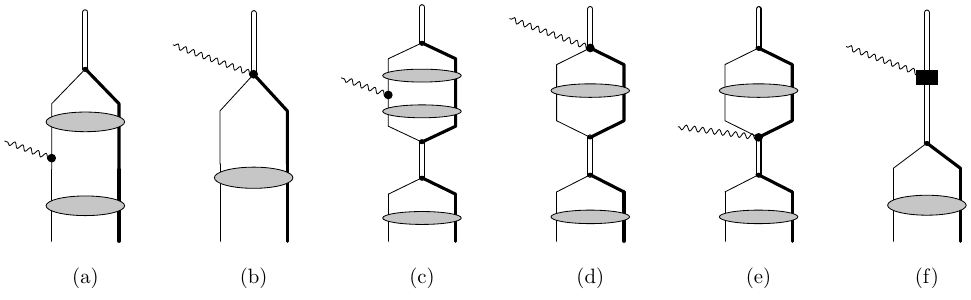}
\caption{Diagrams for $p+\Cp\to\Nt+\gamma$. Thin and thick lines denote the
proton and $\Cp$, double lines the dressed $\tfrac12^+$ and $\tfrac12^-$
dicluster propagators, shaded ovals Coulomb interactions, and wavy lines the
photon. Panels (a) and (b) give external one-body capture and its fixed
seagull. Panels (c) and (d) are the corresponding initial-state-interaction
(ISI) terms. Panel (e) is the corresponding entrance-channel seagull, which
vanishes for the $s$-wave vertex. Panel (f) is the $E1$ counter-term, which
gauge invariance does not fix. In (a) and (c), the photon can couple to either charged particle.}
\label{fig:diagrams}
\end{figure*}

The diagrams contributing to the $s$-wave amplitude are shown in
Fig.~\ref{fig:diagrams}. In panels (a) and (c) the photon couples to either
charged cluster, while (b) and (d) are the seagulls generated by gauging the
$p$-wave final-state vertex. The corresponding seagull from the
entrance-channel vertex, panel (e), vanishes because the $s$-wave vertex
contains no derivative. Panel (f) is the $E1$ counter-term, which gauge
invariance does not fix. We denote the gauge-dictated part of (a)--(e) by
$\mathcal A_{\rm fix}$ and the contribution of (f) by $\mathcal A_{\rm ct}$.

The radial wave functions of the initial $\tfrac12^+$ and the final bound
$\tfrac12^-$ states are
\be
u_0(r)&=&\frac{\cos\delta_0\,F_0(\eta,pr)+\sin\delta_0\,G_0(\eta,pr)}{p},
\nonumber\\
u_1(r) &=&C\,W_{-\eta_B,3/2}(2\gamma r),
\label{eq:u0}
\ee
where $F_\ell(\eta,\rho)$ and $G_\ell(\eta,\rho)$ are the regular and
irregular Coulomb functions~\cite{dlmf}, $W_{\kappa,\mu}$ is the Whittaker
function, and $\delta_0$ is the $s$-wave nuclear phase shift. The constant
$C$ is the ANC of the ground state. The quantities $\delta_0$ and $C$ are
fixed in Sec.~\ref{sec:input}. The resonance dependence of
$\mathcal A_{\rm fix}$ then enters through $\delta_0$.

In the zero-range theory the bound-state wave function has the Whittaker form
for all $r>0$. The region where the physical wave function departs from it is
not resolved, and it enters the amplitude only through the counter-term
operator of Eq.~(\ref{eq:LE1}) and its higher-order counterparts. It is a premise of
low-energy EFTs that short-distance contributions can be replaced by such
local operators. The premise has to be tested numerically by comparing the
predictions with the data order by order, which is the main part of this work
(Sec.~\ref{sec:results}).

As derived in Appendix~\ref{app:e1}, the amplitude $\mathcal A(E)$ reads
\begin{equation}
\begin{split}
\mathcal A=\frac{2\sqrt\pi}{3}\,\Zeff\,\omega\,e^{i(\sigma_0+\delta_0)}
\bigg[&\int_0^\infty\!\!dr\;u_1(r)\,r\,u_0(r)\\
&-\frac{3\pi\,C}{\mu^2\gamma\,\Gm}\,
\frac{\sin\delta_0}{p\,\Chat_0}\,c(E)\bigg],
\end{split}
\label{eq:ampW}
\end{equation}
Here 
$\Zeff\equiv (Z_pm_c-Z_cm_p)/{M}=0.4578$
is the  effective $E1$ charge,
$\sigma_0=\arg\Gamma(1+i\eta)$ is the $s$-wave Coulomb phase,
and
$\Chat_0(\eta)$ is the $s$-wave Coulomb factor 
defined in
Eq.~(\ref{eq:Cl}).
The counter-term $E1$
strength $c(E)$ is given by Eq.~(\ref{eq:cE}). 
The first term
in Eq.~(\ref{eq:ampW}) is $\mathcal A_{\rm fix}$, with the entrance-channel
Coulomb distortion treated exactly and no adjustable electromagnetic
parameter, and the second is $\mathcal A_{\rm ct}$. Both are proportional to
the ANC $C$.

The bracket in Eq.~(\ref{eq:ampW}) is real.
Reversing the sign of the real
bracket leaves the cross section unchanged and generates a
discrete mirror solution for the counter-term amplitude. We return to this
ambiguity in Sec.~\ref{sec:cv}.

We omit finite-wavelength corrections, which enter at higher order in
$\omega$.

\subsection{Hindrance of the $E1$ transition}
\label{sec:hindrance}

The numerical importance of the counter-term is unusually large
in this reaction. If it is switched off, $c(E)=0$, the fitted strong
amplitude gives a radiative width of about $69$~eV for the
$\tfrac12^+$ resonance, compared with the measured $0.49(3)$~eV
\cite{skowronski2023} ($0.098$ Weisskopf units). The observed hindrance therefore appears in the
EFT as a destructive cancellation between $\mathcal A_{\rm fix}$ and
$\mathcal A_{\rm ct}$. To display the cancellation independently of their
common prefactors, we divide the real bracket of Eq.~(\ref{eq:ampW}) by the
coefficient of $c(E)$,
$-3\pi C\sin\delta_0/(\mu^2\gamma\Gm p\Chat_0)$. For the adopted fit,
the two contributions are then, in fm$^{-1}$,
\begin{center}
\begin{tabular}{cccc}
\toprule
$E$ (MeV) & fixed & counter-term & sum \\
\midrule
0.025 & $-3.132$ & $+2.760$ & $-0.372$ \\
0.426 & $-2.963$ & $+2.716$ & $-0.247$ \\
0.700 & $-2.855$ & $+2.686$ & $-0.170$ \\
\bottomrule
\end{tabular}
\end{center}
Near the resonance the net amplitude is about an order of magnitude
smaller than either contribution separately. Defining
$\varepsilon_c\equiv|\mathcal A_{\rm fix}+\mathcal A_{\rm ct}|/
|\mathcal A_{\rm ct}|$, we find $\varepsilon_c=0.091$ there. A given
fractional change in either large contribution therefore produces a fractional
change in the net amplitude about $1/\varepsilon_c\simeq11$ times as large,
and about twice that in the cross section. At $25$~keV the corresponding
amplification factor is $7.4$. We determine the resonance parameters, the
ANC and the counter-term current together in the capture fit.

\section{Parameters of the calculation}
\label{sec:input}

The amplitude of Sec.~\ref{sec:amplitude} depends on three quantities that
remain to be specified, the $s$-wave phase shift $\delta_0$, the ground-state
ANC $C$, and the energy dependence of the counter-term function $c(E)$.

\subsection{Effective-range parametrization}
\label{sec:elastic}

For a channel of orbital angular momentum $\ell$, we combine the
Coulomb-dressed dicluster propagator $D_x$ with its two vertices and
the partial-wave projection and define
\begin{equation}
\begin{split}
\hat D_x(E)&\equiv \frac{\ell!}{(2\ell+1)!!}\, g_x^2D_x(E)
=-\frac{2\pi}{\mu}\,\frac{1}{\Dcal_\ell(E)},\\
\Dcal_\ell(E)&\equiv f_\ell(k)-2\kC h_\ell(\eta),
\end{split}
\label{eq:Dhat}
\end{equation}
where $k=\sqrt{2\mu E}$ and $\eta=\kC/k$. The factor
$\ell!/(2\ell+1)!!$ follows from the vertex normalization of
Appendix~\ref{app:spin}. The
Coulomb-modified effective-range function is
\begin{equation}
f_\ell(k)=k^{2\ell+1}\Cl^2(\eta)\left(\cot\delta_\ell-i\right)+2\kC h_\ell(\eta),
\label{eq:ere}
\end{equation}
where $\delta_\ell$ is the nuclear phase shift. We use
\begin{equation}
\Cl(\eta)=\frac{|\Gamma(\ell+1+i\eta)|\,e^{-\pi\eta/2}}{\Gamma(\ell+1)}
\label{eq:Cl}
\end{equation}
and
\begin{equation}
h_\ell(\eta)=k^{2\ell}\frac{\Cl^2(\eta)}{\Chat_0^2(\eta)}
\left[\psi(i\eta)+\frac{1}{2i\eta}-\log(i\eta)\right],
\label{eq:hl}
\end{equation}
with $\psi$ the digamma function. These definitions follow the conventions
of Ref.~\cite{in2024}. The rescaled Coulomb factor $\Cl$ tends to one for
$\eta\to0$, and the imaginary terms in Eq.~(\ref{eq:ere}) cancel so that
$f_\ell$ is real for real $k$. As in Ref.~\cite{in2024}, we expand
$f_\ell$ about a point near the pole rather than about threshold,
\begin{equation}
f_\ell(k)=\tfrac12 r'_\ell\,(k^2-k_r^2)
-\tfrac14 P'_\ell (k^2-k_r^2)^2+\cdots\,.
\label{eq:pole-ere}
\end{equation}

\subsubsection{The $\tfrac12^+$ channel}
\label{sec:swave}

We parametrize the $\tfrac12^+$ channel by the resonance energy $E_R$ and
the proton width $\Gamma_p$, determined in the capture fit.

We define $E_R$ by $\Re\Dcal_0(E_R)=0$, equivalently
$\delta_0(E_R)=\pi/2$. The width is defined by the phase-shift slope near
the resonance,
\[
\cot\delta_0(E)=\frac{2(E_R-E)}{\Gamma_p}
+\mathcal{O}\!\left((E-E_R)^2\right).
\]
Using $\Re\Dcal_0=k\Chat_0^2(\eta)\cot\delta_0$ with
$k_R=\sqrt{2\mu E_R}$ and $\eta_R=\kC/k_R$, we obtain
\begin{equation}
\Gamma_p=\frac{2k_R\,\Chat_0^2(\eta_R)}
   {\left[2\kC\,\Re\,\dfrac{\mathrm{d}h_0}{\mathrm{d}E}
          -\dfrac{\mathrm{d}f_0}{\mathrm{d}E}\right]_{E=E_R}}.
\label{eq:gamp}
\end{equation}
At LO, $P'_0=0$ and $\mathrm{d}f_0/\mathrm{d}E=\mu r'_0$. The resonance
conditions then give
\begin{equation}
\begin{split}
r'_0&=\frac{1}{\mu}\left[-\frac{2k_R\Chat_0^2(\eta_R)}{\Gamma_p}
      +2\kC\,\Re\left.\frac{\mathrm{d}h_0}{\mathrm{d}E}\right|_{E_R}\right],\\
k_r^2&=k_R^2-\frac{4\kC\,\Re h_0(E_R)}{r'_0}.
\end{split}
\label{eq:swclosed}
\end{equation}
When $P'_0$ is included, we solve the same two resonance conditions
numerically for $(k_r^2,r'_0)$ at each value of $P'_0$.
Equation~(\ref{eq:swclosed}) is recovered as $P'_0\to0$. As an independent
check, we also evaluate the width from
$\Gamma_p=2(d\delta_0/dE)^{-1}$ at $\delta_0=\pi/2$.

We constrain $E_R$ and $\Gamma_p$ with measurements that do not come from the
capture data we fit. That excludes the Felsenkeller
determination~\cite{skowronski2023}, which is extracted from those data, and
evaluated compilations, which draw on the same measurements.

Two measurements remain~\cite{kettner2023,csedreki2023}. After conversion to
the center of mass, Ref.~\cite{kettner2023} gives $E_R=424.65(46)$~keV and
$\Gamma_p=34.0(2)$~keV, its width being an $R$-matrix partial width and
therefore already a center-of-mass quantity.
Reference~\cite{csedreki2023} gives $E_R=424.18(74)$~keV and a laboratory
total width, which becomes $\Gamma_p=35.24(46)$~keV with
$\Gamma_{\rm tot}\simeq\Gamma_p$, the radiative width being negligible.
Their weighted means are
$E_R=424.52\pm0.39$~keV and $\Gamma_p=34.20\pm0.45$~keV, which we use as
the constraint. The two energy measurements agree within $0.5$ times their
combined uncertainty. The two
width measurements differ by $2.5$ times their combined uncertainty, so we
inflate the uncertainty of their weighted mean by the corresponding scale
factor.

\subsubsection{The $\tfrac12^-$ ground state and the ANC}
\label{sec:anc}

For the bound $\tfrac12^-$ channel, the charged $p$-wave pole-residue
relation
\cite{Ryb2014,ryberg2014,zhang2018,higa2018,premarathna2020,RYBERG2016}
relates the ANC of the radial wave function 
$u_1(r)$ in
Eq.~(\ref{eq:u0}) to the residue of the dressed propagator,
\begin{equation}
\Csq=-\frac{2\mu\gamma^2\,\Gm^2}
{\left.\mathrm{d}\Dcal_1/\mathrm{d}E\right|_{E_B}},
\label{eq:anc}
\end{equation}
where $E_B=-S_p$. In the limit $P'_1\to0$,
$\mathrm{d}f_1/\mathrm{d}E=\mu r'_1$, and hence
\begin{equation}
r'_1=\frac{1}{\mu}\left[2\kC\left.\frac{\mathrm{d}h_1}{\mathrm{d}E}
\right|_{E_B}+\left.\frac{\mathrm{d}\Dcal_1}{\mathrm{d}E}\right|_{E_B}\right].
\label{eq:gsclosed}
\end{equation}
The pole condition $\Dcal_1(E_B)=0$ fixes the $p$-wave expansion point.
Equations~(\ref{eq:anc}) and~(\ref{eq:gsclosed}) then trade $r'_1$ for the
ANC. We determine the
ANC from the capture fit. The elastic analysis of Ref.~\cite{in2024} leaves
the $p_{1/2}$ channel poorly constrained, with an uncertainty of about
$13$~MeV in its expansion-point energy.

Reconstructing the truncated $p$-wave effective-range expansion in this way produces a second
pole at $\kappa_{\rm sp}=72.9$~MeV, corresponding to
$E_{\rm sp}=-\kappa_{\rm sp}^2/(2\mu)=-3.07$~MeV. It lies deeper than the
physical bound state at $-S_p=-1.94$~MeV. The capture amplitude depends on
the final-state channel only through $\gamma$ and $C$, so this second pole
does not enter the calculation. 

\subsection{Higher orders}
\label{sec:orders}

With the proton, the $\Cp$ ground state, and the retained $\Nt$
states as explicit fields, Fig.~\ref{fig:diagrams} shows every diagram that
can contribute to the reaction. For the $s_{1/2}\to p_{1/2}$ $E1$ amplitude,
higher orders modify panels (a)--(e) through higher-derivative
interactions and add higher-order counter-term operators to panel (f).

Gauge invariance fixes the seagulls generated by gauging the
higher-derivative vertices. These corrections enter through the
initial $s$-wave effective-range function.

Panel (f) describes the counter-term for the $E1$
transition to the ground state of $\Nt$, with strength $c(E)$ of
Eq.~(\ref{eq:cE}). NLO adds $c_1$ and $P'_0$.
At the following order the short-distance expansion adds $c_2$, while the
strong expansion adds $+Q'_0(k^2-k_r^2)^3$ to $f_0$. We use this counting
in the order-by-order fits of Sec.~\ref{sec:results}.

\section{Results}
\label{sec:results}

\subsection{Data, uncertainties and fit procedure}
\label{sec:fitproc}

We fit 146 $S$-factor points below $E=0.95$~MeV. They consist of 18 points
from the LUNA BGO activation measurement and 32 from the LUNA HPGe
prompt-$\gamma$ measurement~\cite{luna2023}
($67.9\le E\le362.1$~keV), 75 from the Felsenkeller
measurement~\cite{skowronski2023} ($325.6\le E\le620.6$~keV), and 21
activation points of Gy\"urky \emph{et al.}~\cite{gyurky2023}
($271.8\le E\le920.3$~keV). The fit parameters are the counter-term
coefficients of Eq.~(\ref{eq:cE}), the ANC $\Csq$, the $\tfrac12^+$ shape
term $P'_0$ from NLO on, the two resonance parameters, and a normalization
factor $n_k$ for each data set $k$.
The resonance parameters are constrained by the
independent measurements discussed in Sec.~\ref{sec:swave}. All parameters are
refitted at each order.

The fit minimizes
\begin{equation}
\begin{split}
\chi^2=&\sum_k\sum_{i\in k}
\frac{\left[S^{\rm th}(E_i)-n_kS_i\right]^2}
     {(n_k\delta S_i)^2+\left(n_kS_i\,k_i/\Lambda_\chi\right)^2}\\
&+\sum_k\left(\frac{n_k-1}{\Delta_k}\right)^2
\;+\;\chi^2_{\rm con}.
\end{split}
\label{eq:chi2}
\end{equation}
Here $n_k$ is the normalization of data set $k$ and $\Delta_k$ is its
reported systematic uncertainty, as listed in Table~\ref{tab:errors}.
The measured values and their pointwise uncertainties are $S_i$ and
$\delta S_i$, $k_i=\sqrt{2\mu E_i}$, and $\Lambda_\chi$ is the fixed
dispersion scale. Up to NLO we use
\begin{equation}
\chi^2_{\rm con}=
\left(\frac{E_R-424.52}{0.39}\right)^2
+\left(\frac{\Gamma_p-34.20}{0.45}\right)^2
+(P'_0\Lambda^3)^2,
\label{eq:chicon}
\end{equation}
with $E_R$ and $\Gamma_p$ in keV. The last term imposes a naturalness
condition, that $P'_0$ be of order $1/\Lambda^3=11.4$~fm$^3$. It enters at
NLO and acts as a penalty rather than a bound. At the NLO minimum, $\chi^2_{\rm con}\simeq6.2$, while
$\chi^2_{\rm data}=282.3$. The normalization penalties, which contribute
$2.7$, are included separately in Eq.~(\ref{eq:chi2}).

The second term in the denominator of Eq.~(\ref{eq:chi2}) adds a fixed
dispersion. We take the prescription $k_i/\Lambda_\chi$ and the value
$\Lambda_\chi=1$~GeV from our elastic-scattering analysis~\cite{in2024}, and
we do not adjust them here.

The experiments divide their systematic uncertainties into correlated and
pointwise parts in different ways, so for consistency we use each published
total systematic uncertainty $\Delta_k$ as the normalization constraint
(Table~\ref{tab:errors})~\cite{luna2023,skowronski2023,gyurky2023}.

The reported pointwise errors are small, typically $0.3\%$ to $1.2\%$. Without the additional dispersion the fit gives
$\chi^2_{\rm data}/\nu=62.8$, and the discrepancy is not confined to a few
outliers. Of the 75 Felsenkeller points, 41 lie more than $3\sigma$ from the
fit, and the pattern cannot be removed by changing the capture amplitude.
Converting each residual into the energy shift that would remove it gives a
median of $0.6$~keV for Felsenkeller, against $1.9$ to $4.3$~keV for the
other three sets, so the resonance-region data are especially sensitive to
sub-keV changes in energy. We do not fit an energy shift. With this $k_i/\Lambda_\chi$ dispersion,
the relative extra uncertainty is $0.7\%$ at 25~keV, $2.7\%$ at the
resonance, and $4.1\%$ at 0.95~MeV. The adopted fit then gives
$\chi^2_{\rm data}/\nu=282.3/136=2.08$. Contributions from the individual
data sets are listed in Table~\ref{tab:errors}.

\begin{table}[t]
\centering
\caption{Normalization constraints and fitted NLO factors in
Eq.~(\ref{eq:chi2}). The quoted total systematic uncertainty $\Delta_k$
constrains $n_k$. The parenthetical uncertainty on $n_k$ is obtained from
the full NLO covariance. The last column gives the fixed extra dispersion
$k_i/\Lambda_\chi$ at the lowest and highest energies of each data set.
Here $\chi^2_k$ is the contribution of data set $k$ to $\chi^2_{\rm data}$
and $N_k$ is its number of points.}
\label{tab:errors}
\footnotesize
\setlength{\tabcolsep}{3.5pt}
\begin{tabular}{lcccc}
\toprule
data set & $\Delta_k$ & $n_k$ & $k_i/\Lambda_\chi$ (\%) & $\chi^2_k/N_k$ \\
\midrule
LUNA BGO            & $7.9\%$  & $1.011(39)$ & $1.08$--$2.49$ & $1.48$ \\
LUNA HPGe           & $6.9\%$  & $1.012(39)$ & $1.15$--$2.50$ & $0.87$ \\
Felsenkeller        & $8.5\%$  & $1.078(42)$ & $2.37$--$3.28$ & $1.62$ \\
Gy\"urky activation & $7.7\%$  & $0.895(35)$ & $2.17$--$3.99$ & $5.08$ \\
\bottomrule
\end{tabular}
\end{table}

Changing $\Lambda_\chi$ changes the relative weight of the high-energy data.
Lower values give them less weight and also reduce $\chi^2_{\rm data}/\nu$.
The latter reaches unity near $\Lambda_\chi\simeq0.7$~GeV, where
$S(25~\mathrm{keV})$ differs from the adopted result by only $0.3\%$.
Over $\Lambda_\chi=500$--$1500$~MeV, $S(25~\mathrm{keV})$ varies by
$0.7\%$, compared with its $4.1\%$ fit uncertainty, while $\Csq$ stays
between $5.98$ and $6.40$~fm$^{-1}$. Replacing
$k_i/\Lambda_\chi$ by a constant extra dispersion for each data set,
chosen so that $\chi^2_k\simeq N_k$, gives $2.9\%$, $2.0\%$, $3.5\%$,
and $8.0\%$. This changes $S(25~\mathrm{keV})$ by $0.74\%$. We use the
larger of these variations in the error-model entry of
Table~\ref{tab:budget}.

\subsection{Order-by-order fits}
\label{sec:cv}

\begin{figure}[t]
\centering
\includegraphics[width=\columnwidth]{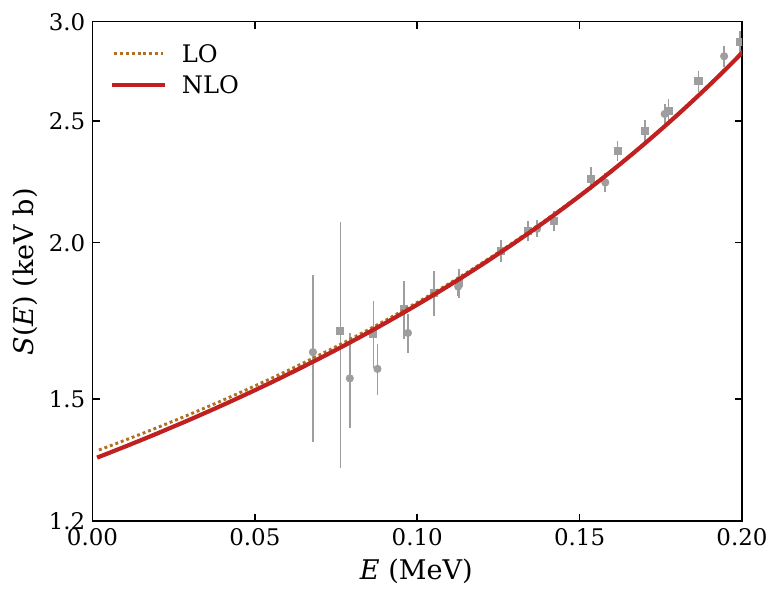}
\caption{$S$ factor in the extrapolation region. The lowest-energy data are shown
in gray with the uncertainties of Eq.~(\ref{eq:chi2}). Every parameter is
refitted at each order as in Table~\ref{tab:orders}.}
\label{fig:orders}
\end{figure}

Table~\ref{tab:orders} lists the LO and NLO fits. All parameters are
refitted at each order. The resonance parameters are nearly unchanged,
with $E_R=424.66$--$424.68$~keV and
$\Gamma_p=35.30$--$35.45$~keV. The independent resonance constraints of
Sec.~\ref{sec:swave} are imposed at both orders. From LO to NLO,
$S(25~\mathrm{keV})$ changes by $-1.0\%$ and $S(0)$ by $-1.3\%$.
The leading counter-term coefficient changes by $4.2\%$, from $2.651$ to
$2.763$~fm$^{-1}$. These shifts are smaller than the nominal per-order
suppression $E/\Etwo\simeq0.1$ at the resonance. The corresponding
low-energy extrapolations are shown in Fig.~\ref{fig:orders}.

The two counter-term coefficients are strongly anticorrelated,
$r(c_0,c_1)=-0.96$, and are also correlated with $\Csq$ and $P'_0$. Their
combination $c(E)$ is better determined than either coefficient separately,
and $|c_1E/c_0|$ reaches only $3.8\%$ at 0.95~MeV.

\begin{table*}[t]
\centering
\caption{Order-by-order fits, with $\nu=138$ at LO and $136$ at NLO. Every
parameter is refitted at each order. All quoted uncertainties are marginal
and are obtained from the full covariance at that order. The coefficients $\mathrm{MeV}^nc_n$ have units
of fm$^{-1}$ and refer to the subtraction of Eq.~(\ref{eq:cdiv}).
An absent term is fixed to zero.}
\label{tab:orders}
\footnotesize
\setlength{\tabcolsep}{3.5pt}
\begin{tabular}{lcccccccc}
\toprule
order & $\chi^2_{\rm data}/\nu$ & $E_R$ (keV) & $\Gamma_p$ (keV)
& $\mathrm{MeV}^0c_0$ & $\mathrm{MeV}^1c_1$
& $P'_0\Lambda^3$ & $C^2$ (fm$^{-1}$)
& $S(0)$ / $S(25)$ (keV\,b) \\
\midrule
LO      & $2.207$ & $424.66(9)$ & $35.45(15)$ & $2.651(5)$ & $0$ & $0$
        & $3.87(18)$ & $1.361(54)$ / $1.441(57)$ \\
NLO     & $2.076$ & $424.68(10)$ & $35.30(17)$
        & $2.763(18)$ & $-0.110(18)$
        & $0.242(70)$ & $6.21(59)$
        & $1.343(56)$ / $1.427(58)$ \\
\bottomrule
\end{tabular}
\end{table*}

\begin{table}[t]
\centering
\caption{Uncertainty budget for the NLO $S$ factor, in keV\,b, with
contributions combined in quadrature.}
\label{tab:budget}
\small
\setlength{\tabcolsep}{4pt}
\begin{tabular}{lcc}
\toprule
source & $\delta S(0)$ & $\delta S(25)$ \\
\midrule
fit covariance            & 0.056 & 0.058 \\
omitted terms             & 0.032 & 0.022 \\
error model (dispersion)  & 0.010 & 0.011 \\
mirror branch                       & 0.023 & 0.017 \\
\midrule
total                     & 0.069 & 0.066 \\
\bottomrule
\end{tabular}
\end{table}

We do not quote next-to-next-to-leading order as a fit result, for the
following reasons. Adding $c_2$ and the next
shape parameter $Q'_0$ lowers $\chi^2_{\rm data}$ by only $1.7$ for two
additional parameters, so that $\chi^2_{\rm data}/\nu$ rises from $2.08$ to
$2.09$. The new terms are constrained mainly by the highest-energy data,
where the present calculation is least complete. Both the omitted
$p$-wave $M1$ capture and the uncalculated nuclear distortion of the
$d_{3/2}$ contribution grow toward the upper end of the fitted interval
(Appendix~\ref{app:other}). Extending only the $s$-wave effective-range and
short-distance expansions would therefore not constitute a complete
calculation at that order. The two constants lower $S(25~\mathrm{keV})$ by
$1.5\%$ and $S(0)$ by $2.3\%$ relative to NLO. These preliminary tests
therefore put the omitted higher-order correction at $0.02$ to
$0.03$~keV\,b in magnitude, which is what enters the budget of
Table~\ref{tab:budget}. That budget is dominated by the fit covariance. We
do not use these shifts to infer a convergence rate.

At NLO, $c_1$ differs from zero by $6.1\sigma$ and $P'_0$ by $3.5\sigma$.
Their correlation is $r(c_1,P'_0)=-0.91$. We therefore adopt NLO as the
highest order for which the added parameters are constrained by the data.

Fixing all four normalization factors to unity provides a simple check.
It leaves $S(25~\mathrm{keV})$ at $1.44$~keV\,b, compared with the adopted
$1.43$~keV\,b, while $\chi^2_{\rm data}/\nu$ rises to $713.7/140=5.10$.

The fitted $\Csq=6.21\pm0.59$~fm$^{-1}$ is larger than the transfer value
$2.69\pm0.36$~fm$^{-1}$~\cite{li2010,artemov2022}. Our preliminary tests of the omitted capture
contributions show that the fitted ANC is much more model dependent than the
low-energy $S$ factor. Adding an estimate of $p$-wave $M1$ capture, and
adding the $\tfrac32^-$ resonance tail as well, each lower $\Csq$, the two
together by about a quarter. The same tests change $S(25~\mathrm{keV})$ by
at most $0.3\%$. Much of the constraint on $\Csq$ comes from the upper end
of the fitted range, where the omitted contributions also grow, so an
accurate comparison with transfer ANCs requires a calculation that includes
them consistently.

\subsubsection*{The mirror branch}

As noted below Eq.~(\ref{eq:ampW}), changing the sign of the real bracket in
that equation leaves the cross section unchanged and maps
\begin{equation}
c(E)\;\longrightarrow\;\frac{2\mu^2\gamma\,\Gm}{3\pi\,C}\,
\frac{p\,\Chat_0}{\sin\delta_0}\int_0^\infty\!\!dr\;u_1\,r\,u_0
\;-\;c(E).
\label{eq:mirror}
\end{equation}
At fixed $\Csq$
and with an untruncated $c(E)$, this is an exact symmetry of the
cross section even when the $d_{3/2}$ term is included. The map
changes the sign of the $s$-wave amplitude and leaves the $d$-wave amplitude
unchanged. Their interference survives in the angular distribution, so
angular data could distinguish the two branches.

Truncating $c(E)$ breaks this symmetry. At LO the mirror branch gives
$\chi^2_{\rm data}=12757$, compared with $305$ for the branch used here.
At NLO the two branches fit the data nearly equally well, with
$\chi^2_{\rm data}=284.2$ and $282.3$. The mirror branch gives
$S(25~\mathrm{keV})$ higher by $1.2\%$ and $S(0)$ higher by $1.7\%$.
We use the branch that is continuously connected to the successful LO fit
and include the NLO branch difference in the uncertainty budget.

\subsubsection*{Other checks}

Raising the lower limit of the fit from $68$ to $200$~keV, or removing any
one of the four data sets, changes $S(25~\mathrm{keV})$ by at most $4\%$.

The fitted parameters are less robust than the $S$ factor. The Gy\"urky
activation data are the only points above $0.621$~MeV, and removing them
raises the uncertainty in $\Csq$ from $0.59$ to $3.5$~fm$^{-1}$ and lowers
the significance of $P'_0$ from $3.5\sigma$ to $0.2\sigma$.

\begin{figure*}[t]
\centering
\includegraphics[width=0.92\textwidth]{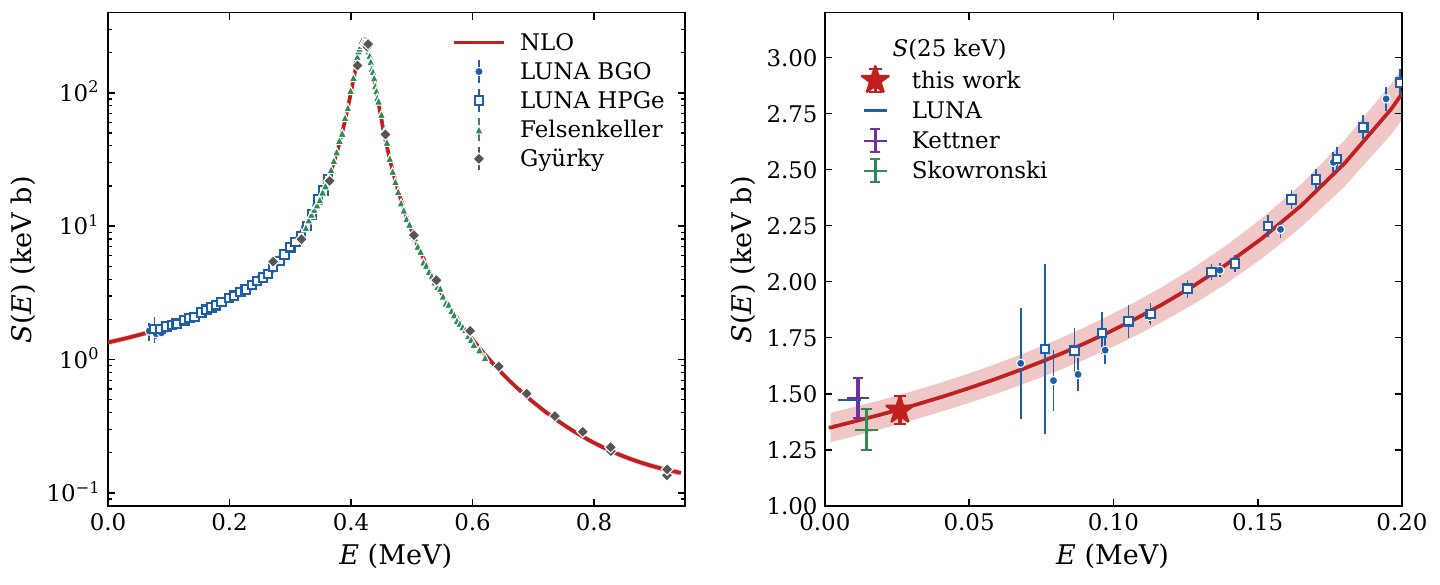}
\caption{Astrophysical $S$ factor. The fitted region is shown in the left
panel. The right panel shows the low-energy extrapolation together with the
present value and three recent extrapolations from LUNA~\cite{luna2023},
Kettner~\cite{kettner2023}, and Skowronski~\cite{skowronski2023}. The latter
two include their quoted uncertainties. The solid red curve is the adopted NLO result and the band is the total
uncertainty of Table~\ref{tab:budget}. The data are multiplied by their
fitted normalizations and shown with the uncertainties of
Eq.~(\ref{eq:chi2}).}
\label{fig:sfactor}
\end{figure*}

These checks do not determine the nuclear distortion of the $d_{3/2}$
capture amplitude at high energy. We have not calculated that correction
in the region where the $d_{3/2}$ contribution becomes dominant.

\subsection{The $S$ factor}
\label{sec:sfac}

Figure~\ref{fig:sfactor} shows the fitted $S$ factor and its low-energy
extrapolation. A cubic extrapolation of the amplitude from $2$--$60$~keV
gives
\begin{equation}
\begin{split}
S(0)&=1.34\pm0.07~\mathrm{keV\,b},\\
S(25~\mathrm{keV})&=1.43\pm0.07~\mathrm{keV\,b}.
\end{split}
\end{equation}

Table~\ref{tab:budget} collects the uncertainty budget. The central values
are $S(0)=1.343$ and $S(25)=1.427$~keV\,b. The fit contribution uses the
full covariance of all ten NLO parameters. The omitted terms are dominated
by the next-order shifts described above, with the omitted $d_{3/2}$
distortion and $p$-wave $M1$ capture adding $0.3\%$ and $0.2\%$. The
error-model entry covers changes of $\Lambda_\chi$ and of the form used for
the additional dispersion, and we take the larger of the two. The mirror
entry is the difference between the two solutions. We add it in quadrature
with the rest as a conservative estimate.

At the adopted NLO minimum, fixing the four normalization factors at their
fitted values reduces the marginal uncertainty in $S(25~\mathrm{keV})$
from $4.08\%$ to $1.13\%$. Thus the normalization factors account for
$92\%$ of the fit variance in $S(25~\mathrm{keV})$ and $90\%$ in $S(0)$.
The low-energy uncertainty is therefore dominated by the experimental
normalizations rather than by the statistical precision of individual
points.

The fitted amplitude also gives the radiative width of the $\tfrac12^+$
state. The measured radiative width is not used in the fit. Since the
resonance-region capture data largely determine the same amplitude, this is
a consistency check rather than an independent prediction. Using only the
resonant part of the calculated cross section in the Breit--Wigner relation
at the peak,
$\Gamma_\gamma=k_R^2\Gamma_p\,\sigma_{\rm res}(E_R)/(4\pi)$, gives
$\Gamma_\gamma=0.470$~eV. Using the full amplitude at the peak in the same
conversion gives $0.467$~eV. Both agree with the measured $0.49(3)$~eV
within $0.8\sigma$.

\begin{figure*}[t]
\centering
\includegraphics[width=0.92\textwidth]{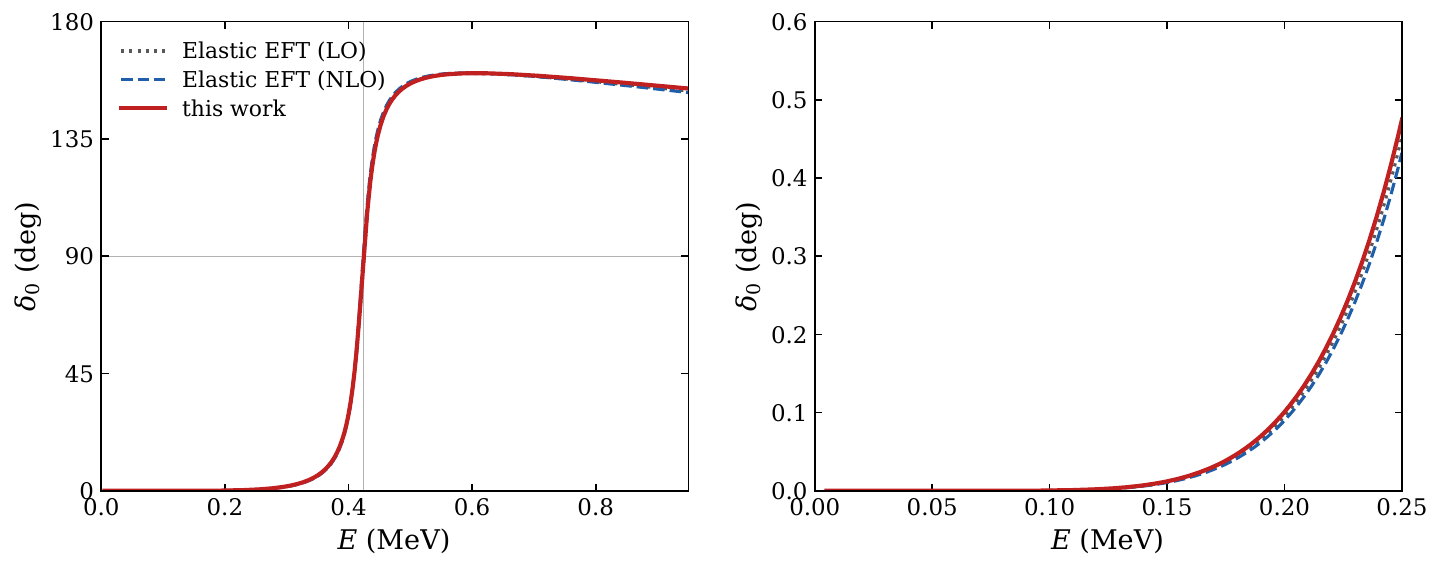}
\caption{$s$-wave $p$-$\Cp$ phase shift from the elastic-scattering
parameters of Ref.~\cite{in2024}, with LO shown as a dotted line and NLO
as a dashed line, together with the parameters determined here (solid).
The left panel shows the full fit window with $E_R$ marked. The right panel
shows the region below the resonance, where the three curves differ by at
most $0.011^\circ$ up to $0.2$~MeV in absolute terms, corresponding
to about $11\%$ of $\delta_0$ itself.}
\label{fig:elastic}
\end{figure*}

\subsection{Comparison with previous results}
\label{sec:cmp}

Our $S(25~\mathrm{keV})=1.43\pm0.07$~keV\,b agrees within uncertainties
with the recent $R$-matrix extrapolations of Kettner \emph{et al.},
$1.48\pm0.09$~keV\,b~\cite{kettner2023}, and Skowronski \emph{et al.},
$1.34\pm0.09$~keV\,b~\cite{skowronski2023}. The LUNA low-temperature
rate~\cite{luna2023} corresponds to about $1.47$~keV\,b. These values are
shown in Fig.~\ref{fig:sfactor}. Our
$S(0)=1.34\pm0.07$~keV\,b is $1.1\sigma$ below the Solar Fusion~III
recommendation $1.44\pm0.06$~keV\,b~\cite{solarfusion3}. The two
potential-model calculations of 2025~\cite{tursunov2025,dubovichenko2025}
give $S(0)$ between $1.33$ and $1.48$~keV\,b.

The fitted proton width is $\Gamma_p=35.30\pm0.17$~keV. It agrees with the
center-of-mass value $35.24(46)$~keV of Ref.~\cite{csedreki2023} and is
higher than the $34.0(2)$~keV result of Ref.~\cite{kettner2023}. The fitted
value lies $2.4$ standard deviations above the constraint mean
$34.20$~keV and has a smaller marginal uncertainty. Removing the resonance
constraint altogether raises $\Gamma_p$ to $35.48$~keV and changes
$S(25~\mathrm{keV})$ by $0.02\%$, so the capture data rather than the
constraint fix the width.

The $s$-wave phase shift obtained here differs from the NLO elastic fit of
Ref.~\cite{in2024} by $7$--$13\%$ relative to $\delta_0$ over
$0.15$--$0.30$~MeV, as shown in Fig.~\ref{fig:elastic}. The relative
difference is amplified because $\delta_0$ is small below the resonance.
In absolute terms the curves differ by at most $0.011^\circ$ up to
$0.2$~MeV. We obtain $P'_0\Lambda^3=0.242(70)$, or
$P'_0=2.75(80)$~fm$^3$, while the NLO elastic analysis gives
$-2.55(11)$~fm$^3$. The two solutions also differ in $r'_0$, $1.591$
against $1.447$~fm. At the
same $E_R$ and $\Gamma_p$, the two values of $P'_0$ change $\delta_0$ by
$0.007^\circ$ at $0.2$~MeV and by $2.5^\circ$ at $0.95$~MeV.


\section{Reaction rate}
\label{sec:rate}

\begin{figure}[t]
\centering
\includegraphics[width=\columnwidth]{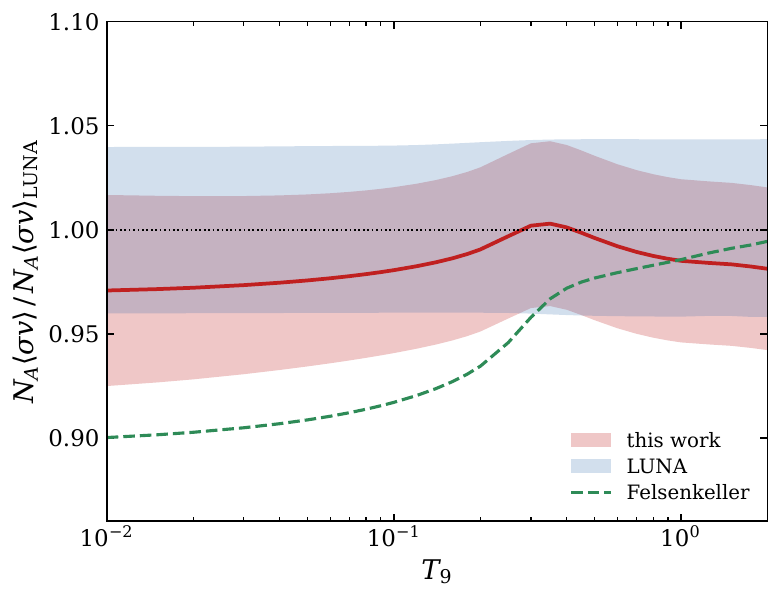}
\caption{Reaction rate relative to the LUNA determination~\cite{luna2023}.
The solid red curve is the present result and the red band is the total
uncertainty from Table~\ref{tab:budget}, evaluated at each temperature.
The blue band shows the LUNA uncertainty and the dashed green curve the
Felsenkeller rate~\cite{skowronski2023}.}
\label{fig:rate}
\end{figure}

The thermonuclear rate is
\begin{equation}
\begin{split}
N_A\langle\sigma v\rangle = &3.7320\times10^{10}\,\mu_u^{-1/2}T_9^{-3/2}\\
&\times\int_0^\infty\! S(E)\,e^{-2\pi\eta-11.6045E/T_9}\,dE,
\end{split}
\end{equation}
in cm$^3$mol$^{-1}$s$^{-1}$ for $S$ in MeV\,b and $E$ in MeV.
Here $\mu_u=0.929254$ is the reduced mass in atomic mass units and $T_9$
is the temperature in GK. Table~\ref{tab:rate} lists selected values, and
Fig.~\ref{fig:rate} compares the result with the LUNA rate. Over
$0.01\le T_9\le2$, the ratio of the present rate to the LUNA rate lies
between $0.971$ and $1.003$. The two rates therefore differ by no more than
$2.9\%$ over this interval.

We propagate the sources in Table~\ref{tab:budget} through the rate integral
at each temperature. The resulting uncertainty ranges from about $4.0\%$ at
$T_9\simeq0.35$ to $4.7\%$ at $T_9=0.01$. We evaluate the integral up to
$E=1.8$~MeV. The part above the fitted range, $E>0.95$~MeV, contributes only
$0.3\%$ of the rate at $T_9=2$ and is negligible below $T_9=1$. We therefore
quote the rate only for $T_9\le2$. At higher temperatures the omitted
$p$-wave $M1$ capture becomes increasingly important and should be included.

\begin{table*}[t]
\centering
\caption{Reaction rate in cm$^3$mol$^{-1}$s$^{-1}$. Here $E_0$ is the Gamow
peak energy. The central column is the adopted rate. The low and high columns
are obtained by combining the contributions of Table~\ref{tab:budget} in
quadrature at each temperature. Valid for $T_9\le2$.}
\label{tab:rate}
\small
\begin{tabular}{cccccc}
\toprule
$T_9$ & $E_0$ (keV) & low & central & high & LUNA~\cite{luna2023} \\
\midrule
0.01 &  18.2 & $1.073\times10^{-19}$ & $1.126\times10^{-19}$ & $1.179\times10^{-19}$ & $1.160\times10^{-19}$\\
0.02 &  29.0 & $3.445\times10^{-14}$ & $3.608\times10^{-14}$ & $3.772\times10^{-14}$ & $3.712\times10^{-14}$\\
0.05 &  53.4 & $1.179\times10^{-8}$ & $1.231\times10^{-8}$ & $1.283\times10^{-8}$ & $1.262\times10^{-8}$\\
0.10 &  84.7 & $1.788\times10^{-5}$ & $1.864\times10^{-5}$ & $1.940\times10^{-5}$ & $1.901\times10^{-5}$\\
0.20 & 134.4 & $6.528\times10^{-3}$ & $6.798\times10^{-3}$ & $7.069\times10^{-3}$ & $6.863\times10^{-3}$\\
0.50 & 247.6 & $1.380\times10^{1}$ & $1.437\times10^{1}$ & $1.494\times10^{1}$ & $1.443\times10^{1}$\\
1.00 & 393.1 & $5.203\times10^{2}$ & $5.419\times10^{2}$ & $5.634\times10^{2}$ & $5.501\times10^{2}$\\
2.00 & 624.0 & $2.137\times10^{3}$ & $2.225\times10^{3}$ & $2.314\times10^{3}$ & $2.268\times10^{3}$\\
\bottomrule
\end{tabular}
\end{table*}

\section{Summary and outlook}
\label{sec:summary}

We have calculated ground-state $\Cp(p,\gamma)\Nt$ capture in cluster
effective field theory through NLO. The low-energy reaction is dominated by
$s_{1/2}\to p_{1/2}$ $E1$ capture. Its amplitude is strongly hindered by
destructive interference between the fixed contribution and the
counter-term $E1$ current. At NLO the fit adds the counter-term slope
$c_1$ and the $\tfrac12^+$ shape term $P'_0$, both favored by the full data
set.
Our adopted result is
\begin{equation*}
\begin{split}
S(0)&=1.34\pm0.07~\mathrm{keV\,b},\\
S(25~\mathrm{keV})&=1.43\pm0.07~\mathrm{keV\,b}.
\end{split}
\end{equation*}
The value at 25~keV agrees with the recent $R$-matrix extrapolations, while
$S(0)$ is $1.1\sigma$ below the Solar Fusion~III recommendation.

The capture data also constrain the $\tfrac12^+$ resonance strongly.
Removing the resonance constraint changes $\Gamma_p$ from $35.30$ to
$35.48$~keV and changes $S(25~\mathrm{keV})$ by only $0.02\%$. The
radiative width obtained from the fitted capture amplitude is $0.470$~eV,
consistent with the measured $0.49(3)$~eV within $0.7\sigma$.

We use the fixed momentum-dependent dispersion prescription of our earlier
elastic analysis and obtain $\chi^2_{\rm data}/\nu=2.08$ without rescaling
the covariance. We estimate the truncation uncertainty from the change
produced by adding $c_2$ and $Q'_0$. The total uncertainty in the low-energy
$S$ factor is dominated by the experimental normalization factors.

Our preliminary tests change $\Csq$ substantially while changing
$S(25~\mathrm{keV})$ by at most $0.3\%$, so the fitted
$\Csq=6.21\pm0.59$~fm$^{-1}$ is more model dependent than the low-energy
$S$ factor.

The resulting thermonuclear rate differs from the LUNA rate by at most
$2.9\%$ over $0.01\le T_9\le2$, with an uncertainty of about
$4$--$5\%$. A preliminary calculation with the finite-wavelength one-body
currents and the proton magnetization changes $S(E)$ by less than $0.03\%$
after refitting. The remaining
$p$-wave $M1$ contribution and higher multipoles, together with the nuclear
distortion of direct $d_{3/2}$ capture, become more important toward the
upper end of the fitted range and at higher temperatures. A more complete calculation of these
terms is needed before extending the rate above $T_9=2$. On the experimental
side, angular distributions would help resolve the mirror ambiguity through
$s$--$d$ interference.

\section*{Data availability}

The data that support the findings of this article are openly available in
Refs.~\cite{luna2023,skowronski2023,gyurky2023}.

\section*{Acknowledgments}
We thank Prof. S.-W. Hong for useful discussions.
The work of TSP was supported by the IBS grant funded by the Korea government
(No. IBS-R031-D1). The work of YHS was supported by the Rare Isotope Science
Project of the Institute for Basic Science (IBS-I001-01).

\appendix

\section{Spin and vertex conventions}
\label{app:spin}

The $\Cp$ core has spin zero, so the channel spin is the proton spin,
$s=\tfrac12$. The channels are therefore labeled by the orbital angular
momentum $\ell$ and the total angular momentum
$j=\ell\pm\tfrac12$. We use spherical derivatives
$\nabla_{\pm1}=\mp(\nabla_x\pm i\nabla_y)/\sqrt2$ and
$\nabla_0=\nabla_z$, and define the relative derivative by
\begin{equation}
\psi_p\,i\overleftrightarrow{\nabla}\,\psi_c
\equiv
-\frac{1}{M}\,\psi_p\left(
m_c\,i\overleftarrow{D}-m_p\,i\overrightarrow{D}
\right)\psi_c ,
\label{eq:reld}
\end{equation}
with the covariant derivative of Eq.~(\ref{eq:lag}) and the arrows
indicating the field each one acts on. On proton and core plane waves it
gives $(m_c\bm p_p-m_p\bm p_c)/M=\bm q$, so $i\overleftrightarrow{\nabla}$ is
the coordinate-space form of the relative momentum used in the scattering
amplitude.

For the channels retained in this work, the vertices are
\begin{equation}
\begin{split}
P_{\frac12^+,s}(\psi_p,\psi_c)
&=\psi_{p,s}\,\psi_c,\\
P_{\frac12^-,s}(\psi_p,\psi_c)
&=\sum_{m,s'} C^{\frac12 s}_{1m,\frac12 s'}\,
\psi_{p,s'}\,i\overleftrightarrow{\nabla}_m\,\psi_c,\\
P_{\frac32^-,s}(\psi_p,\psi_c)
&=\sum_{m,s'} C^{\frac32 s}_{1m,\frac12 s'}\,
\psi_{p,s'}\,i\overleftrightarrow{\nabla}_m\,\psi_c,\\
P_{\frac52^+,s}(\psi_p,\psi_c)
&=\sum_{m_1,m_2,m,s'}
C^{2m}_{1m_1,1m_2}
C^{\frac52 s}_{2m,\frac12 s'}\\
&\qquad\times
\psi_{p,s'}\,
i\overleftrightarrow{\nabla}_{m_1}\,
i\overleftrightarrow{\nabla}_{m_2}\,
\psi_c .
\end{split}
\label{eq:Pxapp}
\end{equation}

\section{$E1$ amplitudes in the long-wavelength limit}
\label{app:e1}

Here we collect the formulas for the $E1$ amplitude of the
$s_{1/2}\to p_{1/2}$ transition in the long-wavelength limit, where the
photon exponentials in the one-body convection current are replaced by one
and the charge weights add to $\Zeff$. It is convenient to define the
velocity integral
\begin{equation}
\mathcal V[\phi_0]=\int_{r_{\min}}^\infty\!\!dr\;
u_1\Big(\phi_0'-\frac{\phi_0}{r}\Big),
\label{eq:Vfun}
\end{equation}
where $\phi_0$ and $u_1(r)=C\,w_1(r)$, with
$w_1(r)=W_{-\eta_B,3/2}(2\gamma r)$, are the $s$- and $p$-wave radial
functions and $r_{\min}$ is a coordinate-space cutoff, taken to zero once
the divergence is absorbed into the counter-term.

Near the origin the two functions read
\begin{equation}
\begin{split}
u_1&\rightarrow\frac{a}{r}\left(1-\kC r\right)+\mathcal O(r),\\
\phi_0&\rightarrow b_0\Big(1+2\kC r\ln\frac{r}{r_{\rm sub}}\Big)+b_1r
+\mathcal O(r^2\ln r),
\end{split}
\label{eq:origin}
\end{equation}
with $a=C/(\gamma\Gm)$. In the zero-range theory $u_1$ is the irregular
solution of the $p$-wave equation, while $\phi_0$ carries both the regular
and the irregular $s$-wave behavior. Here $r_{\rm sub}$ is an arbitrary
positive length whose dependence is absorbed into $b_1$, so that the
results do not depend on it.

Using the radial equations $\phi_0''=(2\kC/r-p^2)\phi_0$ and
$u_1''=(2\kC/r+2/r^2+\gamma^2)u_1$ together with $p^2+\gamma^2=2\mu\omega$,
the velocity integral can be rewritten as
\begin{equation}
\begin{split}
\mathcal V[\phi_0]=&\;\mu\omega\!\int_{r_{\min}}^\infty\!\!dr\;r\,u_1\phi_0
-\frac{ab_0}{r_{\min}}\\
&-3a\kC b_0\ln\frac{r_{\min}}{r_{\rm sub}}
-\frac32\,ab_1-\frac12\,a\kC b_0 ,
\end{split}
\label{eq:Vparts}
\end{equation}
up to terms of order $r_{\min}\ln r_{\min}$. The seagull generated by gauging
the derivative in the $p$-wave vertex is
\begin{equation}
\mathcal V_{\rm surf}[\phi_0]
=-\frac{3}{2}\,u_1(r_{\min})\,\phi_0(r_{\min}),
\label{eq:Bfun}
\end{equation}
so that in the difference
$\overline{\mathcal V}\equiv\mathcal V-\mathcal V_{\rm surf}$ the logarithm
and $b_1$ cancel, leaving
\begin{equation}
\overline{\mathcal V}[\phi_0]=
\mu\omega\!\int_{r_{\min}}^\infty\!\!dr\;r\,u_1\phi_0
+\left(\frac{1}{2r_{\min}}-2\kC\right)a\,b_0 .
\label{eq:Vbar}
\end{equation}
Only this combination enters the $E1$ amplitude.

The initial state is $\phi_0=e^{i\delta_0}u_0$ with $u_0$ of
Eq.~(\ref{eq:u0}), which separates into
\begin{equation}
\phi_0=\frac{F_0(\eta,pr)+e^{i\delta_0}\sin\delta_0\,H^+_0(\eta,pr)}{p} ,
\label{eq:phisplit}
\end{equation}
where $H^+_0=G_0+iF_0$. With $\mathcal K\equiv(2\sqrt\pi/3)(\Zeff/\mu)\,e^{i\sigma_0}$, the diagrams
of Fig.~\ref{fig:diagrams} give
\begin{equation}
\begin{aligned}
\mathcal A_{\rm (a)}+\mathcal A_{\rm (b)}&=
\frac{\mathcal K}{p}\,\overline{\mathcal V}[F_0],\\
\mathcal A_{\rm (c)}+\mathcal A_{\rm (d)}&=
\frac{\mathcal K}{p}\,e^{i\delta_0}\sin\delta_0\,
\overline{\mathcal V}[H^+_0],\\
\mathcal A_{\rm (e)}&=0,
\end{aligned}
\label{eq:pairs}
\end{equation}
where in each pair the photon attaches to a charged cluster in the first
diagram and gives $\mathcal V$, and to the $p$-wave vertex in the second and
gives $-\mathcal V_{\rm surf}$. The sum of the above becomes even simpler,
\begin{equation}
\mathcal A_{\rm (a)}+\cdots+\mathcal A_{\rm (e)}
=\mathcal K\,\overline{\mathcal V}[\phi_0]
=\mathcal A_{\rm fix}+\mathcal A_{\rm div},
\label{eq:Asum}
\end{equation}
with
\begin{equation}
\begin{aligned}
\mathcal A_{\rm fix}&=\frac{2\sqrt\pi}{3}\,\Zeff\,\omega\,
e^{i(\sigma_0+\delta_0)}\!\int_0^\infty\!\!dr\;u_1\,r\,u_0,\\
\mathcal A_{\rm div}&=\mathcal K\Big(\frac{1}{2r_{\min}}-2\kC\Big)a\,b_0,
\end{aligned}
\label{eq:diagrams}
\end{equation}
where $\mathcal A_{\rm fix}$ is the dipole matrix element between $u_1$ and
$u_0$~\cite{siegert1937,friar1984}, and
\begin{equation}
a\,b_0=\lim_{r\to0}r\,u_1\phi_0
=\frac{C}{\gamma\Gm}\,\frac{e^{i\delta_0}\sin\delta_0}{p\,\Chat_0} .
\label{eq:ab0}
\end{equation}
Here $\mathcal A_{\rm (a)}+\mathcal A_{\rm (b)}$ is finite, while
$\mathcal A_{\rm (c)}+\mathcal A_{\rm (d)}$ is not, and
$\mathcal A_{\rm div}$ is the part of it that comes from the origin.
Finally, diagram (f) reads
\begin{equation}
\mathcal A_{\rm (f)}=-\mathcal K\,\frac{3\pi}{\mu}\,
\omega\,a\,b_0\,\mathring c(E).
\label{eq:Af}
\end{equation}
We set the divergent part of $\mathring c(E)$, the $c_{\rm div}$ of
Eqs.~(\ref{eq:cren}) and (\ref{eq:cdiv}), to cancel $\mathcal A_{\rm div}$
exactly, so that
$\mathcal A_{\rm fix}+\mathcal A_{\rm div}+\mathcal A_{\rm (f)}$ reduces to
Eq.~(\ref{eq:ampW}) of the main text with the renormalized $c(E)$.

At 25~keV, $96.3\%$ of the external dipole integral
$\int_0^\infty dr\,w_1(r)\,r\,F_0(\eta,\rho)$ comes from
$r>1/\Lambda=2.25$~fm, and its median radius is $10.8$~fm. This is the
peripherality quoted in Sec.~\ref{sec:counting}.


\section{Capture from other partial waves}
\label{app:other}

The shifts from the $d_{3/2}$ strong interaction and from $p$-wave $M1$
capture are estimated here and included in the uncertainty budget.

For direct $d_{3/2}\to p_{1/2}$ $E1$ capture, no new low-energy constant
is needed at the order considered here. Using the same bound-state wave
function as in the $s$-wave amplitude and the regular Coulomb function
$F_2$, the long-wavelength result can be written as
\begin{equation}
\begin{split}
\sigma_d&=2\left(\frac{\mathcal I_2(E)}{\mathcal I_0(E)}\right)^2
\sigma_{\rm ext},\\
\mathcal I_\ell(E)&\equiv\int_0^\infty\!\!dr\;w_1(r)\,r\,F_\ell(\eta,\rho),
\end{split}
\label{eq:dwave}
\end{equation}
where $\sigma_{\rm ext}$ is the long-wavelength cross section from the
external $s$-wave term alone. It is obtained from Eq.~(\ref{eq:ampW}) by
setting $c(E)=0$ and $\delta_0=0$, with $\mathcal I_0$ replacing the full radial
integral. The numerator and denominator in Eq.~(\ref{eq:dwave}) are then
evaluated with the same $E1$ operator.

The factor $2$ is the ratio of reduced matrix elements,
\begin{equation}
\frac{
\left|\langle p_{1/2}\|Y_1\|d_{3/2}\rangle\right|^2
}{
\left|\langle p_{1/2}\|Y_1\|s_{1/2}\rangle\right|^2
}=\frac{1/\pi}{1/(2\pi)}=2.
\label{eq:y1ratio}
\end{equation}
All remaining factors cancel in the ratio of
Eq.~(\ref{eq:dwave}). After angular integration the $s$- and $d$-wave
contributions do not interfere, so $\sigma_d$ is added directly to the
cross section. The angular distribution does contain their interference,
as noted in Sec.~\ref{sec:cv}.

At this order the strong interaction in the $d_{3/2}$ entrance channel is
dropped and only direct Coulomb capture is kept, so no gauged $d$-wave
vertex appears. A short-distance $E1$ operator for this transition
carries two more derivatives than the $s$-wave counter-term operator and lies
beyond the order retained here.

We estimated the effect of the omitted $d_{3/2}$ nuclear interaction by
giving its phase shift the hard-sphere value for a radius $1/\Lambda$, and
again with that phase shift multiplied by 25. After refitting, these
variations change $S(25~\mathrm{keV})$ by at most $0.3\%$. The test bounds
the low-energy extrapolation. It is weaker near $0.8$--$0.92$~MeV, where
the direct $d_{3/2}$ contribution is large and where the fitted ANC is most
sensitive to the high-energy data.

We also estimated one-body $M1$ capture, both with and without the
$\tfrac32^-$ resonance tail. Only the $p_{3/2}$ entrance wave contributes.
For $p_{1/2}$ the leading long-wavelength matrix element vanishes, since
within that channel the magnetic operator is proportional to $\bm J$ and
the bound and scattering radial states are orthogonal. After refitting, the
$p_{3/2}$ estimates change $S(25~\mathrm{keV})$ by at most $0.2\%$, while
the fitted $\Csq$ moves appreciably. This sensitivity is one reason the
fitted ANC is more model dependent than the low-energy $S$ factor.

Both the $d_{3/2}$ distortion and the estimated $p$-wave $M1$ contribution
grow toward the upper end of the fitted interval. The tests use one-body
electromagnetic operators alone, without the corresponding seagull
currents. We therefore take the resulting changes, $0.3\%$ and $0.2\%$ in
$S(25~\mathrm{keV})$, as the omitted $d_{3/2}$ and $p$-wave $M1$ entries of
Table~\ref{tab:budget}, and no more than that. A consistent treatment,
together with the relevant seagull currents and higher multipoles, is left
for a future calculation.

\end{document}